# Selective multiple domain wall injection using spin-orbit torque

Ziyan Luo, Wugang Liao, Yumeng Yang, Chunxiang Zhu and Yihong Wu[a]

*Department of Electrical and Computer Engineering, National University of Singapore, 4*

*Engineering Drive 3, Singapore 117583, Singapore*

We demonstrate from both simulation and experiment a simple scheme for selective injection of multiple domain walls in a magnetic nanowire. The structure consists of a side-contact misaligned Hall bar made of ferromagnet/heavy metal bilayers. The combination of current-induced spin-orbit torque and an external magnetic field allows for the formation of localized domains with specific magnetization direction and length, thereby creating domain walls in predetermined locations. With the side contacts at two sides misaligned for a distance that is comparable to the contact width, it is possible to create densely packed domains by simply applying current between different pairs of side contacts. Simulation results show that the proposed scheme is scalable to a large number of domains with its dimension limited only by the domain wall width.

---

[a] Author to whom correspondence should be addressed: elewuyh@nus.edu.sg

Magnetic domain walls (DW) have enormous promise in both memory and logic applications, particularly the DW in magnetic heterostructures with perpendicular magnetic anisotropy (PMA) as it can be driven to move at fast speed by both a magnetic field and in-plane current[1-3]. Unlike single-domain devices such as magnetic random access memory, the DW-based devices typically require a string of well-defined domain walls. Therefore, the key to realizing such kind of devices is controlled injection and motion of domain walls along a pre-defined track, *e.g.*, magnetic nanowires. Conventionally, DWs are injected by local magnetic field generating from current-carrying ($I_{inj}$) wire together with an external magnetic field, and moved with a shift pulse current $I_{shift}$ [4-8], as illustrated in Fig. 1(a). Other methods such as manipulation of the shape and dimension of the magnetic wire[9,10] and modification of the anisotropy locally using focused ion beam (FIB) irradiation[11,12] were also reported. In all these designs, typically an externally applied assistant field is required to create domain walls with desired magnetization configurations. Recently, Phung *et al.*[13] have demonstrated a highly efficient in-line DW injector as illustrated schematically in Fig. 1(b), where a 90º magnetization boundary created by FIB irradiation in a nanowire with PMA is used to inject DWs without the need of any localized Oersted or external field. This greatly simplifies the design of DW-based devices.

In all the aforementioned approaches, however, the DW is injected in a one-by-one fashion. After a DW is injected, it has to be shifted away from the nucleation region so as to allow for the injection of next DW. This kind of bit-wise injection or writing does not only affect the operation speed but also makes it difficult to control the spacing between neighboring DWs precisely. Here, we demonstrate a scheme which allows for injection of multiple DWs with well-defined spacing in a "block-writing" manner. In addition, it will also shorten the writing time for large amount of data as domain wall motion is only required after a block of DWs is written instead of a single DW as in the conventional approaches. As a proof-of-concept design, Fig. 1(c) shows the geometry of the proposed structure.

Central to this scheme is a side-contact misaligned Hall (SMH) bar composed of a heavy metal (HM) and a ferromagnetic (FM) strip or nanowire with PMA; the latter is deposited directly on the HM Hall bar and is narrower than the HM Hall bar itself. When a charge current flows through different lateral contacts (1a-1d, 2a-2c) of the Hall bar, a localized current is generated in a well-defined region in the FM/HM bilayer, thereby allowing for selective switching of magnetization locally. The underlying mechanism is the spin-orbit torque (SOT) induced by an in-plane charge current in FM/HM heterostructures. Although the exact mechanism is still being debated, it is generally accepted that two types of torques are present in the FM/HM heterostructures, one is called field-like (FL) and the other is (anti)damping-like (DL). Phenomenology, the two types of torques can be modelled by $\vec{T}_{DL} = \tau_{DL}\vec{m} \times [\vec{m} \times (\vec{j} \times \vec{z})]$ and $\vec{T}_{FL} = \tau_{FL}\vec{m} \times (\vec{j} \times \vec{z})$, respectively, where $\vec{m}$ is the magnetization direction, $\vec{j}$ is the in-plane current density, $\vec{z}$ is the interface normal, and $\tau_{FL}$ and $\tau_{DL}$ are the magnitudes of the FL and DL torques, respectively[14-16]. The DL torque offers an efficient mechanism for switching the magnetization of FM with PMA, as already demonstrated in numerous works recently[14-27], though an externally applied field in the current direction is typically required to achieve definite switching. The polarity of switching is determined by both the current and assistant field directions, as demonstrated previously in single element devices. We show that, by controlling the current direction using the side-contact misaligned Hall bars, it is possible to inject multiple magnetic domain selectively with precisely determined lengths. In addition to the side-contact spacing and width, the domain length can also be controlled by the current density. The smallest domain length is limited either by lithography or the domain wall width, whichever is larger.

Experiments have been carried out to verify the selective writing mechanism based on the SMH design. A large SMH structure, with a central area of 10 μm (width) × 100 μm (length) and side contacts of 10 μm (width) × 30 μm (length), were first fabricated using combined techniques of liftoff and sputtering deposition. The Hall bar stack, comprising of Ta(1.5)/Pt(3)/Co(0.6)/Pt(1), was deposited on SiO$_2$/Si substrates by magnetron sputtering with a base pressure of $2 \times 10^{-8}$ Torr and working pressure of $3 \times 10^{-3}$ Torr. The number inside the parentheses is thickness in *nm*. The second step was to selectively remove the top Co(0.6)/Pt(1) layers to form a 500 nm strip of Co(0.6)/Pt(1) in the central region of the Hall device by electron beam lithography and ion milling. This is to improve the current distribution in the region that is directly underneath the Co layer. Both the side and longitudinal electrodes were formed by Ta(5)/Cu(200)/Pt(10) multilayers. Current induced switching experiments were performed at room temperature in combination with a scanning magneto-optic Kerr effect (SMOKE) microscope setup. The latter was set in a polar mode to capture the magnetization distribution after each switching operation.

In the first set of experiments, we examine how the length of the reversed domain depends on the current density. This is important because, unlike the case of current applied in the longitudinal direction, the current density is non-uniform when the current is applied across the mutually shifted side contacts. To this end, we supply a current pulse with a duration of 5 ms and current density (*j*) in the range of 1.05 – 1.27 $\times 10^{12}$ A/m$^2$ between the side electrodes 1a and 2a of a upward saturated Hall bar. The cross-section is estimated via a combination of bottom Ta(1.5)/Pt(3) layer with a width of 10 μm and upper Co(0.6)/Pt(1) layer with a width of 500 nm. Note that this is just an estimation as the current density itself is non-uniform. A field of 650 Oe is applied in *x*-direction, *i.e.*, the wire direction, to assist deterministic switching. Fig. 2(a) illustrates the Hall bar superimposed with the saturated domain-free background image. The narrow brighter region at the center is the magnetic wire, and *w* is the width of both the Hall bar and the side-contact, as well as the spacing between the side contacts. To facilitate discussion, we divide each side-contact into two halves by the dashed-lines and the magnetic segments between two adjacent dashed-lines are labelled as no. 1 - 4 (circled). As the writing is done by manual wire connection, it usually takes about 10 sec to complete the single domain injection while up to 30 sec for the injection of multiple domains. During this process, the stage would drift, resulting in blurring of the SMOKE image. To avoid this and keep the results consistent, we adopted the following steps to write both single and multiple domains: i) saturating the entire Hall bar in upward direction with a positive pulse current under an assistant field, ii) writing the segment(s) with negative pulse current with the same external field, iii) refocusing on the sample surface to capture the SMOKE image of written segment(s), iv) saturating the entire Hall bar in upward direction again and capturing the background image, and v) subtracting the image taken in iii) from background image to obtain the differential SMOKE image. From the captured SMOKE image, we obtained the length of reversed region ($d$) and its dependence on current density is shown in Fig. 2(b). Reversible switching of each segment is achieved by changing the polarity of electrical bias across different pairs of electrodes, *e.g.*, 1a-2a for segment no.1. Shown alongside the figure at right-hand side are the corresponding SMOKE images captured at different current densities (indicated by the dashed line). To ensure consistency, after each writing/capturing cycle, the sample is re-set into saturation state before changing the current and capturing another image. As can be seen from the figure, nearly a linear dependence between the domain length and current density is obtained, providing an effective mechanism for controlling the domain length in a deterministic manner. It is worth noting that the smallest domain length is about $0.5w$ and the longest is around $1.2w$. This means that one has a full control of both the polarity and length of the injected domain by simply changing the current direction and density. Qualitatively the results can be understood as follows. By taking into account the non-uniformity of current distribution in the *xy*-plane (see Fig.1(c) for the definition of coordinate axes), the damping-like effective field can be written as $\vec{m} \times (\vec{j} \times \vec{z}) = (j_x m_z, j_y m_z, -j_x m_x - j_y m_y)$. When the field applied in *x*-direction is sufficiently strong, one can assume $j_y = 0$, therefore, the switching of magnetization from up to down direction or vice versa is determined by the sign and magnitude of $j_x m_x$. At a fixed external field, the onset of switching is thus determined by $j_x$ only. Since $j_x$ is non-uniform, the switching will take place in the region where $j_x$ is largest and then gradually expands to two sides with smaller current density. The range of 0.5 $w$ – 1.2 $w$ is obtained under the specific experimental condition. When the current density goes smaller than $1.05 \times 10^{12}$ A/m², there is still switching locally, but the domain becomes discontinuous. On the other hand, in principle, the maximum length can still be increased up to around 2 $w$, we did not do so in order to avoid electric breakdown of the wire. Despite the difference in injection mechanism, it is worth mentioning that the estimated current density here is comparable to the value reported for the injection geometry of Fig. 1(b)[13], which is $2.9 \times 10^{12}$ A/m². Both are 1-2 orders of magnitude smaller than the current density used in conventional method.

To develop a quantitative understanding of the results shown in Fig. 2(b), we simulated the current distribution of the portion of the Hall bar shown in the inset of Fig. 2(c) using COMSOL. For simplicity, we assume that the Hall bar consists of only a single layer with a lateral dimension that is the same of the actual device. The mesh-size used is 20 nm × 20 nm × 1.2 nm, and the number of meshes in thickness direction is 5. The conductivity and thickness used were $3.16 \times 10^6$ S/m (measured value) and 6.1 nm (total thickness of the Hall bar), respectively. Since the SOT mainly originates from the bottom Pt layer, we only analyze the current distribution on $xy$ plane at the middle of Hall bar in thickness direction. Fig. 2c shows the normalized current distribution in $x$ direction ($j_x$) along the central line of the portion of the Hall bar as a function of $x$ position (see inset for the $x$ coordinate). At a bias voltage of $V_0$ = 10 V between electrode 1a and 2a, the maximum current density in $x$-direction is $j_m(V_0) = 1.1 \times 10^{12}$ A/m². Based on the simulated current distribution, a threshold current density ($j_{th}$) of $6.81 \times 10^{11}$ A/m² is required for full switching of the region between side contact 1a and 2a, *i.e.*, segment 1 in Fig. 2a. The length of switched region is $w$. The corresponding colormap of $j_x/j_m(V_0)$ is shown in the inset of Fig.2c. Note that the absolute value is not of the main concern here because we are only interested in how the domain length varies with the current density. To this end, we gradually scale down the bias voltage $V$ such that $j_m(V)/j_m(V_0)$ would scale according to the experimentally values used in Fig. 2(b), and then calculate the domain length $d$ under each bias voltage $V$ by using the same $j_{th}$. Here, $j_m(V)$ is the maximum $j_x$ obtained at a bias of $V$. In this case, we assumed that the region with $j_x > j_{th}$ will be switched. Fig. 2(d) shows the normalized domain length ($d/w$) as a function of normalized maximum current density $j_m(V)/j_m(V_0)$. We can see that the simulation results correlate well with the experimental results in Fig. 2(b). The results suggest that switching is based on the SOT mechanism and the length of switched region can be controlled precisely by the current density.

We now discuss the case of writing multiple segments by applying current between electrodes 1a-2a, 2a-1b, 1b-2b, and 2b-1c, separately. As is with the case of single element, the length of reversed region increases with the bias current, and eventually overlaps with neighboring segments. A more quantitative evaluation of the current density dependence is the percentage of reversed regions ($P_r$) as compared to the physical length between 1a and 1c, as shown in Fig.3. The captured SMOKE images are also shown alongside the figure. Again, a nearly linear dependence is obtained up to 100% reversal, suggesting that densely packed domains can be created using this scheme by simply controlling the current density.

With the confirmation of adequate control over the length of reversed domains, we now turn to selectivity of the writing process. To demonstrate this, we wrote 16 combinations of four magnetic segments between 1a and 1c as shown in Fig. 4(a). To put it into perspective of digital data storage, we defined up (↑) / down (↓) magnetization states as 1/0, *e.g.,* we labeled the magnetic configuration "↓↑↑↑"as "0111". As mentioned above, for any single segment, we are able to write up or down magnetization state by applying current with sufficient strength and appropriate polarity between a pair of electrodes at opposite sides of the Hall bar. For consecutive zeros or ones, however, instead of writing individual segments separately, we can also write them once by applying current across contacts at the same side of the Hall bar. Take "0011" as an example, instead of applying current between 1a-2a, 2a-1b consecutively, we can simply apply current between 1a-1b. Similarly, "0001" and "0000" states can be achieved by applying current between 1a-2b and 1a-1c, respectively. Fig. 4(b)-(e) shows the simulation results of current distribution with the corresponding magnetization states given at the bottom left corner of each panel. The white arrow shows the overall flowing direction of current while the small black arrows indicates the local current direction. The colormap superimposed with the vector plot is the $x$-component of current distribution, with normalized $j_x$ as the color legend. In correspondence with the experimental results, regions with higher intensity of $j_x$ are located between the halving lines of selected electrodes. Once the current is sufficiently large, the region between the halving lines of the two electrodes will be reversed, resulting in a newly written state. Although we only demonstrate selective writing of four segments, it is apparent that it can be readily scaled up to any numbers as desired, at least without any physical limit.

As a proof-of-concept demonstration, we used a relatively large Hall bar and wide wire to facilitate SMOKE imaging. A question to ask naturally is: does the scheme still work when the Hall bar is scaled down to a much smaller dimension? In order to answer this question, we performed the simulation on the same structure as illustrated in the inset of Fig. 2(c), by scaling down both the lateral dimension and applied voltage simultaneously and without changing the thickness, *i.e.*, keeping it at 6.1 nm. Fig.5 shows the normalized $j_x$ distribution as a function of $x$ position for Hall bar with width $w = 10$ nm, 20 nm, 50 nm, 1 μm and 10 μm, respectively. In the plot, $x$ is normalized to the width $w$. Due to finite thickness effect, the normalized $j_x$ distribution along the center line becomes less sharply defined at the edges when $w$ decreases. However, there is still a sufficient margin for current selectivity even when the lateral dimension is reduced to 10 nm. Therefore, the smallest domain length is likely to be determined by the fabrication process or domain wall length, whichever is larger. The domain wall length, in turn, is determined by material parameters, Dzyaloshinskii-Moriya interaction, interface roughness, *etc*. Therefore, the smallest achievable domain length may vary from one system to another, depending on the FM/HM combination. It is worth pointing out that the same densely packed domains cannot be injected using normal Hall bars with aligned side-contacts.

To conclude, we have devised a method for selective injection of domain walls based on the SOT effect. Experimental results together with simulation demonstrate that both the length of injected domain and the gap between adjacent domains can be controlled precisely via the applied current amplitude. By using different combinations of side contacts, domains with predetermined length can be created selectively. In addition, simulation results on structures with different dimension show that the scheme can apply to structures with smaller dimensions without any fundamental limit except for the DW width itself or fabrication processes. Compared to previously reported approaches, our design can potentially lead to more accurate and faster injection of DWs, which will benefit both DW-based memory and logic devices.

This work is supported by the Singapore National Research Foundation, Prime Minister's Office, under its Competitive Research Programme (Grant No. NRF-CRP10-2012-03) and Ministry of Education, Singapore under its Tier 2 Grant (Grant No. MOE2013-T2-2-096). Y.H.W. is a member of the Singapore Spintronics Consortium (SG-SPIN).

**FIGURE CAPTIONS:**

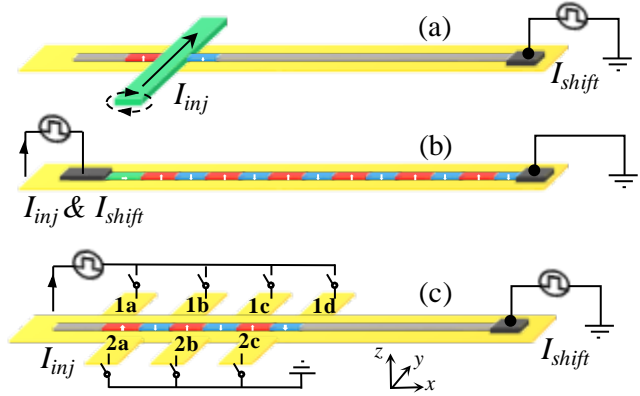

FIG. 1. Schematics of different types of DW injectors. (a) Conventional DW injector comprising of orthogonally arranged magnetic track and a metallic wire for applying writing current; (b) In-line DW injector consisting of magnetic regions whose magnetizations are aligned non-collinearly for facilitating DW injection; (c) SOT-based DW injector using the side-contact misaligned Hall bar structure.

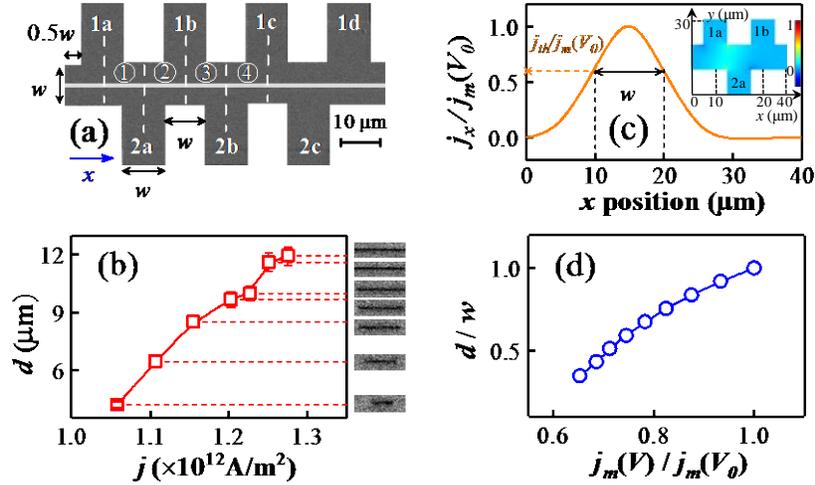

FIG. 2. (a) Side-contact misaligned Hall bar superimposed with saturated SMOKE image. (b) Reversed domain length as a function of pulse current density in segment no.1. The corresponding SMOKE images under different pulse current intensities are shown at the right-hand side of the figure (indicated by dashed line). The error bars indicate the measurement accuracy of domain length. (c) Normalized $j_x$ as a function of $x$ position. The inset is the simulated current distribution along $x$ direction when $d = w$. Origin of $x$-axis is also indicated in the inset, which is the colormap of $j_x$. (d) Normalized domain length, $d/w$ as a function of $j_m(V)/j_m(V_0)$.

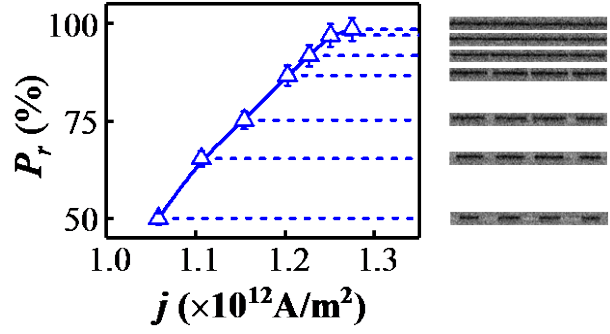

FIG. 3. Percentage of magnetization reversal ($P_r$) in four segments as a function of pulse current density. Shown in the right-hand side are the corresponding SMOKE images at different current densities. The error bars indicate the measurement accuracy of domain length.

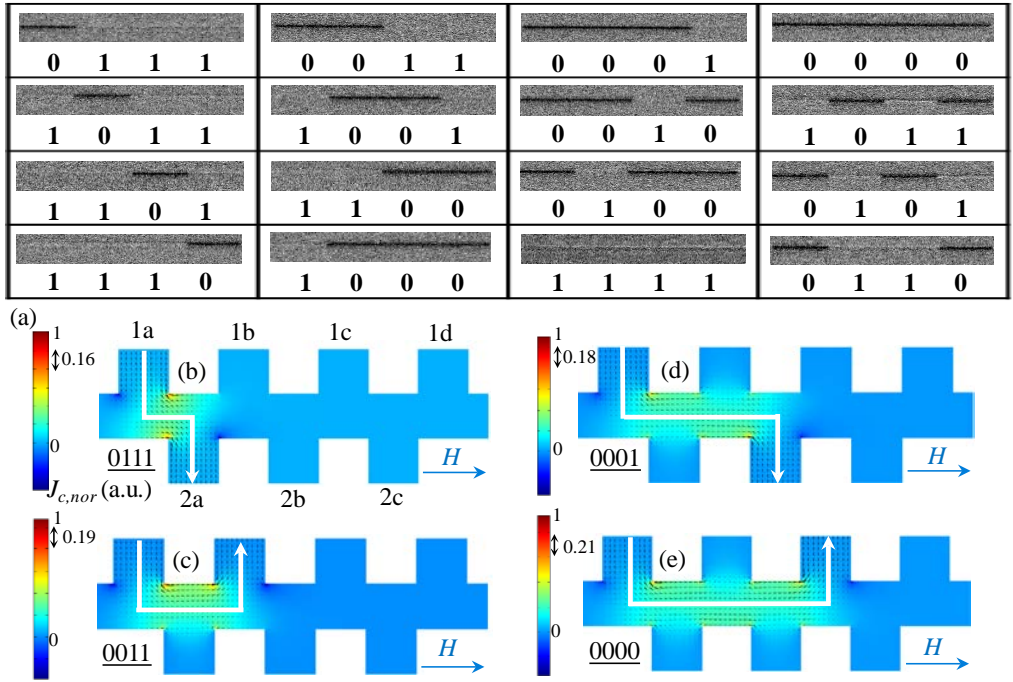

FIG. 4. (a) SMOKE image of 16 combinations of magnetization states between contacts 1a and 1c. "1" and "0" refer to the up and down magnetization states, respectively. (b)-(e) Simulated current distribution in the Hall bar with current running between electrodes 1a-2a, 1a-1b, 1a-2b, and 1a-1c, which are corresponding to the magnetization states of "0111", "0011", "0001", and "0000", respectively. Where applicable, white arrows indicate the overall current flowing direction while the black arrows represent the local current direction. The color-map is the distribution of x-component of the current.

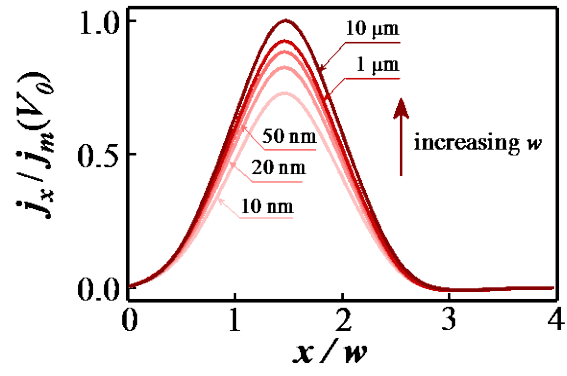

FIG. 5. Normalized $j_x$ as a function of $x$ position with $w$ ranging from 10 nm to 10 μm. $x$ is normalized to $w$.